\documentclass[aps,prd,preprint,showpacs,eqsecnum,nofootinbib]{revtex4}

\usepackage{amssymb,amsmath,feynarts}

\def\bra#1#2{[#1\,#2]}
\def\ket#1#2{\langle #1\,#2\rangle}
\def\fig#1{Fig.~{\ref{#1}}}

\begin{document}

\noindent \hfill Brown-HET-1570

\vskip 1 cm

\title{Three Applications of a Bonus Relation for Gravity Amplitudes}

\author{Marcus Spradlin}

\affiliation{Brown University, Providence, Rhode Island 02912, USA}

\author{Anastasia Volovich}

\affiliation{Brown University, Providence, Rhode Island 02912, USA}

\author{Congkao Wen}

\affiliation{Brown University, Providence, Rhode Island 02912, USA}

\begin{abstract}
Arkani-Hamed et.~al.~have recently shown that all
tree-level scattering amplitudes in maximal supergravity
exhibit exceptionally
soft behavior when two supermomenta are taken to infinity in a particular
complex direction, and that
this behavior implies new non-trivial
relations amongst amplitudes in addition to the well-known
on-shell recursion relations.
We consider the application of these new `bonus relations' to MHV amplitudes,
showing that they can be used quite generally to relate $(n-2)!$-term formulas
typically
obtained from recursion relations to $(n-3)!$-term formulas related to
the original BGK conjecture.
Specifically we provide (1) a direct proof of a 
formula presented
by Elvang and Freedman,
(2) a new formula based on one due to
Bedford et.~al., and (3) an alternate proof of a formula recently obtained
by Mason and Skinner.
Our results also provide the first direct proof that the
conjectured BGK formula, only very recently proven via completely
different methods,
satisfies the on-shell recursion.

\end{abstract}


\maketitle

\section{Introduction}

The enormous progress
we have witnessed over the past few years
in our understanding of the structure and calculability
of gluon scattering amplitudes amply demonstrates that simple
amplitudes do not require simple Lagrangians.
Perhaps counterintuitively,
it has recently been suggested~\cite{ArkaniHamed:2008gz} that the quantum
field theory with the simplest amplitudes
may in fact be
${\mathcal{N}} = 8$
supergravity (SUGRA).

However it is clear that we are still very far from completely exposing
the alleged simplicity of SUGRA amplitudes, even at tree level.
To see this
one need look
no further than the maximally helicity violating (MHV) graviton
amplitudes.
The formula originally conjectured by Berends, Giele and
Kuijf~\cite{Berends:1988zp},
not proven until 20 years later by Mason and Skinner~\cite{Mason:2008ms},
has for $n>4$ particles the
form\footnote{
For simplicity
we avoid committing to a choice of which two particles
$i$ and $j$
have negative helicity. This means that~(\ref{BGK}) and~(\ref{PT})
are to be understood as superspace expressions with the overall
delta-function $\delta^{2 {\mathcal{N}}}(q) = \delta^{2 {\mathcal{N}}}(
\sum \widetilde{\lambda}_i^\alpha \eta_i^A)$ of supermomentum
conservation suppressed.  To restore the helicity information one
would multiply~(\ref{BGK}) by $\ket{i}{j}^8$ 
for negative helicity gravitons,
and~(\ref{PT}) by $\ket{i}{j}^4$ for negative helicity gluons.}
\begin{equation}
\label{BGK}
{\mathcal M}_n =
\sum_{\mathcal{P}(2,\ldots,n-2)}
\frac{\bra{1}{2} \bra{n-2}{n-1}}{\ket{1}{n-1}}
\left(
\prod_{i=1}^{n-3} \prod_{j=i+2}^{n-1} \ket{i}{j}\right)
\prod_{k=3}^{n-3}
[k|p_{k+1} + p_{k+2} + \cdots + p_{n-1} |n\rangle\,,
\end{equation}
where the sum runs over all permutations of the labels
$\{2,\ldots,n-2\}$.  This expression, obtained from the KLT
relations~\cite{Kawai:1985xq} between open and closed
string amplitudes (reviewed in~\cite{Bern:2002kj})
does not seem particularly simple, especially
when contrasted with the remarkable Parke-Taylor
formula for color-ordered MHV gluon
amplitudes~\cite{Parke:1986gb,Berends:1987me}
\begin{equation}
\label{PT}
A(1,2,\ldots,n) = \frac{1}{\ket{1}{2} \ket{2}{3} \cdots \ket{n}{1}}\,.
\end{equation}

A number of authors have observed that some gravity 
amplitudes have exceptionally soft behavior as the momenta of two
particles are taken to infinity in a certain complex
direction\footnote{Moreover, it has been argued~\cite{Bern:2007xj}
that the soft behavior of tree amplitudes is of direct importance
for UV cancellations in SUGRA loop amplitudes~\cite{Bern:2006kd}.}
(see for
example~\cite{Bedford:2005yy,Cachazo:2005ca,BjerrumBohr:2005jr,Benincasa:2007qj,Bern:2007xj}).
One result of~\cite{ArkaniHamed:2008gz}
is the finding that in fact all
SUGRA amplitudes
fall off like $1/z^2$ for large $z$
as the supermomenta of two particles are 
taken to infinity in a particular complex superdirection.
This stands in contrast to SYM,
where amplitudes only fall off like $1/z$
(see~\cite{ArkaniHamed:2008yf, Badger:2008rn} for treatments of
other theories).
The $1/z$ falloff common to SYM and SUGRA
allows one to rewrite the contour integral
$\oint \frac{dz}{z}\, {\mathcal{M}}(z)$
as a sum over residues with no contribution at infinity~\cite{Britto:2005fq},
leading
to the well-known BCF recursion relations~\cite{Britto:2004ap}
for
the physical amplitude ${\mathcal{M}}(0)$ (studied
for gravity in~\cite{Bedford:2005yy,Cachazo:2005ca} and fully proven
first in~\cite{Benincasa:2007qj}).
The $1/z^2$ falloff special to gravity
allows the same to be done for
$\oint dz\, {\mathcal{M}}(z)$,
implying new non-trivial relations between tree amplitudes which
we call the `bonus relations'.

All known $n$-graviton MHV formulas in the
literature~\cite{Berends:1988zp,Nair:2005iv,Bedford:2005yy,Bern:2007xj,Elvang:2007sg,Mason:2008ms}
fall into two categories: those
which come from solving the on-shell recursion
have $(n-2)!$ terms (since the recursion treats two lines as special),
while those obtained from manipulating the BGK formula preserve its
$(n-3)!$ terms.

In this paper we show that the MHV bonus relation
may be used quite generally to relate an $(n-2)!$-term formula
to one with $(n-3)!$-terms that obey a slightly simplified recursion.
Three specific applications are presented:
a proof of a formula proposed by
Elvang and Freedman~\cite{Elvang:2007sg},
a new formula based on one due to Bedford~et.~al.~\cite{Bedford:2005yy},
and an alternate proof of a formula recently obtained by
Mason and Skinner~\cite{Mason:2008ms}.

Since two of the formulas we prove~((\ref{EF2}) and~(\ref{MS})) are
known~\cite{Bern:2007xj,Elvang:2007sg,Mason:2008ms} to be
equivalent to the original BGK conjecture~(\ref{BGK}),
our work provides
as a byproduct
the first direct proof that the BGK formula
(which Mason and Skinner derived using completely different methods)
satisfies the on-shell recursion.

Our work is only a small step in the decades-long march
towards a better understanding of the structure of tree-level graviton
amplitudes.
Over the years a variety of different approaches
have shed light
on this problem in addition to those mentioned above,
including
string-based methods~\cite{Bern:1993wt},
Lagrangian-level manipulations~\cite{Bern:1999ji},
twistor string inspired ideas~\cite{Giombi:2004ix,BjerrumBohr:2006sg}
such as the MHV vertex approach~\cite{BjerrumBohr:2005jr,Bianchi:2008pu},
and exploitation of $E_{7(7)}$
symmetry~\cite{Kallosh:2008ic,ArkaniHamed:2008gz,Kallosh:2008rr,Kallosh:2008ru}.
Despite all of this progress it seems clear that much of the structure
is still elusive (see for example~\cite{Bern:2008qj} for some specific
open questions).
In particular,
in this paper we only consider
MHV amplitudes although the bonus relations have implications
for all amplitudes.
Moreover, in SYM it has been found that
NMHV amplitudes, for example, satisfy certain sum rules~\cite{Kiermaier:2008vz}
which are not a consequence of large $z$ behavior.
If the promise of~\cite{ArkaniHamed:2008gz} is realized then we can
expect an even richer story to emerge for SUGRA.

\section{Cashing in the Bonus for MHV Amplitudes}

\begin{figure}
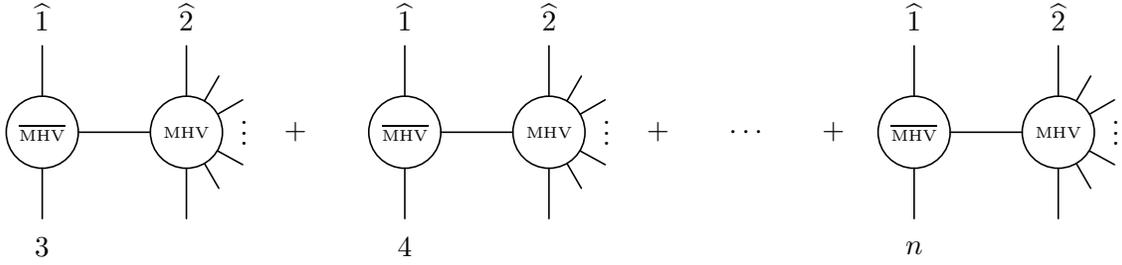

$$
{\hbox{\lower 57.5pt\hbox{
\begin{feynartspicture}(120,120)(1,1)
\FADiagram{} \FAProp(5.,7.5)(5.,7.5)(5.,12.5){/Straight}{0}
\FAProp(16.7678,11.7678)(16.7678,11.7678)(13.2322,8.2322){/Straight}{0}
\FAProp(7.5,10.)(12.5,10.)(0.,){/Straight}{0}
\FAProp(5.,7.5)(5.,4.)(0.,){/Straight}{0}
\FAProp(15.,7.5)(15.,4.)(0.,){/Straight}{0}
\FAProp(5.,12.5)(5.,16.)(0.,){/Straight}{0}
\FAProp(15.,12.5)(15.,16.)(0.,){/Straight}{0} \FALabel(5.,2.)[]{$3$}
\FALabel(5.,18.)[]{$\widehat{1}$}
\FALabel(15.,18.)[]{$\widehat{2}$}
\FALabel(5.,10.)[]{$\overline{\text{\tiny MHV}}$}
\FALabel(15.,10.)[]{\tiny MHV}
\FALabel(19.,10.5)[]{$\vdots$}
\FAProp(17.1651,11.25)(18.8971,12.25)(0.,){/Straight}{0}
\FAProp(16.25, 12.1651)(17.25, 13.8971)(0.,){/Straight}{0}
\FAProp(16.25, 7.8349)(17.25, 6.1029)(0.,){/Straight}{0}
\FAProp(17.1651, 8.7)(18.8971, 7.75)(0.,){/Straight}{0}
\end{feynartspicture}
}}} + {\hbox{\lower 57.5pt\hbox{
\begin{feynartspicture}(120,120)(1,1)
\FADiagram{} \FAProp(5.,7.5)(5.,7.5)(5.,12.5){/Straight}{0}
\FAProp(16.7678,11.7678)(16.7678,11.7678)(13.2322,8.2322){/Straight}{0}
\FAProp(7.5,10.)(12.5,10.)(0.,){/Straight}{0}
\FAProp(5.,7.5)(5.,4.)(0.,){/Straight}{0}
\FAProp(15.,7.5)(15.,4.)(0.,){/Straight}{0}
\FAProp(5.,12.5)(5.,16.)(0.,){/Straight}{0}
\FAProp(15.,12.5)(15.,16.)(0.,){/Straight}{0}
\FALabel(5.,2.)[]{$4$} \FALabel(5.,18.)[]{$\widehat{1}$}
\FALabel(15.,18.)[]{$\widehat{2}$}
\FALabel(5.,10.)[]{$\overline{\text{\tiny MHV}}$}
\FALabel(15.,10.)[]{\tiny MHV}
\FALabel(19.,10.5)[]{$\vdots$}
\FAProp(17.1651,11.25)(18.8971,12.25)(0.,){/Straight}{0}
\FAProp(16.25, 12.1651)(17.25, 13.8971)(0.,){/Straight}{0}
\FAProp(16.25, 7.8349)(17.25, 6.1029)(0.,){/Straight}{0}
\FAProp(17.1651, 8.7)(18.8971, 7.75)(0.,){/Straight}{0}
\end{feynartspicture}
}}} + \qquad \cdots\qquad + {\hbox{\lower 57.5pt\hbox{
\begin{feynartspicture}(120,120)(1,1)
\FADiagram{}
\FAProp(3.,7.5)(3.,7.5)(3.,12.5){/Straight}{0}
\FAProp(14.7678,11.7678)(14.7678,11.7678)(11.2322,8.2322){/Straight}{0}
\FAProp(5.5,10.)(10.5,10.)(0.,){/Straight}{0}
\FAProp(3.,7.5)(3.,4.)(0.,){/Straight}{0}
\FAProp(13.,7.5)(13.,4.)(0.,){/Straight}{0}
\FAProp(3.,12.5)(3.,16.)(0.,){/Straight}{0}
\FAProp(13.,12.5)(13.,16.)(0.,){/Straight}{0}
\FALabel(3.,2.)[]{$n$}
\FALabel(3.,18.)[]{$\widehat{1}$}
\FALabel(13.,18.)[]{$\widehat{2}$}
\FALabel(3.,10.)[]{$\overline{\text{\tiny MHV}}$}
\FALabel(13.,10.)[]{\tiny MHV}
\FALabel(17.,10.5)[]{$\vdots$}
\FAProp(15.1651,11.25)(16.8971,12.25)(0.,){/Straight}{0}
\FAProp(14.25, 12.1651)(15.25, 13.8971)(0.,){/Straight}{0}
\FAProp(14.25, 7.8349)(15.25, 6.1029)(0.,){/Straight}{0}
\FAProp(15.1651, 8.7)(16.8971, 7.75)(0.,){/Straight}{0}
\end{feynartspicture}
}}}
$$\vskip -0.5cm
\caption{
All factorizations contributing to~(\ref{BCF}) or~(\ref{bonus})
for the MHV amplitude ${\mathcal{M}}_n$.}
\label{mhvfig2}
\end{figure}

We follow the conventions of~\cite{ArkaniHamed:2008gz} in
choosing the supersymmetry-preserving
shift~\cite{Brandhuber:2008pf,ArkaniHamed:2008gz}
\begin{align}
\label{bcfshift}
\lambda_{\widehat{1}}(z) &= \lambda_1 + z \lambda_2, \cr
\widetilde{\lambda}_{\widehat{2}}(z) &= \widetilde{\lambda}_2 -
z \widetilde{\lambda}_1, \cr
\eta_{\widehat{1}}(z) &= \eta_1 + z \eta_2\,.
\end{align}
Although bonus relations hold for all amplitudes, their application
is simplest for MHV amplitudes to which
we now restrict our attention.
In this case there is only a single relevant type of factorization,
shown in~\fig{mhvfig2}.
Let us define the subamplitude
\begin{equation}
\label{Mdef}
M_k = \int d^8 \eta\ {\mathcal{M}}_{\text{L}}(\widehat{1},k,-\widehat{P}(z_k))
\frac{1}{(p_1+p_k)^2} {\mathcal{M}}_{\text{R}}
(\widehat{2},3,\ldots,k\!\!\!\!\times,\ldots,n,
\widehat{P}(z_k))
\end{equation}
as the expression corresponding
to the diagram in~\fig{mhvfig2}
with particle $k \in \{ 3,\ldots, n\}$ joining particle $1$
on the left side of the factorization.
Here
\begin{equation}
\widehat{P}(z) = p_1+p_k + z \lambda_k \widetilde{\lambda}_1
\end{equation}
is the shifted intermediate momentum crossing the diagram and
\begin{equation}
z_k = - \frac{\ket{1}{k}}{\ket{2}{k}}
\end{equation}
is the value of $z$ at which $\widehat{P}(z)$ goes on-shell.
In terms of the subamplitudes defined in~(\ref{Mdef})
the BCF recursion is simply
\begin{equation}
\label{BCF}
{\mathcal{M}}_n = M_3 + M_4 + \cdots + M_n\,,
\end{equation}
while the bonus relation takes the form
\begin{equation}
\label{bonus}
0 = z_3 M_3 + z_4 M_4 + \cdots + z_n M_n\,.
\end{equation}
We can use this equation to delete any one $M_k$,
say $M_3$, in~(\ref{BCF}), arriving at
\begin{equation}
\label{newmhv}
{\mathcal{M}}_n = \sum_{k=4}^n \left(1 - \frac{z_k}{z_3}\right) M_k 
= \sum_{k=4}^n \frac{\ket{1}{2} \ket{3}{k}}{\ket{1}{3} \ket{2}{k}}
M_k\,
\end{equation}
with the help of the Schouten identity.
This is almost identical to the BCF recursion~(\ref{BCF}) except
that one buys a reduction
in the number of terms from $n-2$ to $n-3$
at the price of inserting a relatively simple factor into the sum.
When applied recursively
this method reduces the number of
terms
in an $n$-graviton
amplitude
from $(n-2)!$ to $(n-3)!$.

\section{Applications}

\subsection{Direct proof of a formula due to Elvang and Freedman}

In~\cite{Elvang:2007sg} Elvang and Freedman presented
two new formulas for MHV gravity amplitudes in terms of squared
MHV SYM amplitudes.  First, the BCF recursion was used to prove the formula
\begin{equation}
\label{EF1}
{\mathcal{M}}_n = \sum_{{\mathcal{P}}(3,\ldots,n)}
F(1,2,\ldots,n)
\end{equation}
with
\begin{align}
F(1,2,\ldots,n) &= \ket{1}{n} \bra{n}{1} \left(
\prod_{s=4}^{n-1} \beta_s\right)
A(1,2,\ldots,n)^2\,,\cr
\beta_s &= - \frac{\ket{s}{s+1}}{\ket{2}{s+1}}
\langle 2 | p_3 + p_4 + \cdots + p_{s-1} |s]\,.
\end{align}
The second formula
is
\begin{equation}
\label{EF2}
{\mathcal{M}}_n = \sum_{{\mathcal{P}}(4,\ldots,n)}
\frac{\ket{1}{2} \ket{3}{4}}{\ket{1}{3} \ket{2}{4}}
F(1,2,\ldots,n)\,,
\end{equation}
where the distinguished particle 3 can be chosen freely from
the set $\{3,\ldots,n\}$ without changing the result as long
as the sum includes all permutations of the remaining $n-3$ elements.
This formula was obtained by manipulating a slightly reprocessed version
of the BGK formula from~\cite{Bern:2007xj}.
We will now show that~(\ref{EF2}) follows directly
from~(\ref{EF1}) and~(\ref{newmhv}).

The proof proceeds by induction, beginning
with the cases $n=4,5$ already shown to be correct
in~\cite{Elvang:2007sg}.  We
now have to show that~(\ref{EF2}) continues to hold for $n+1$ gravitons
if we allow ourselves to assume that it holds up to and including $n$
gravitons.  Now the factor $F(1,2,\ldots,n)$ was shown in~\cite{Elvang:2007sg}
to satisfy the BCF recursion already, which means that we know it satisfies
\begin{equation}
\label{FBCF}
\int d^8 \eta\ {\mathcal{M}}_{\text{L}}(\widehat{1},n+1,-\widehat{P}_{n+1})
\frac{1}{(p_1+p_{n+1})^2}
F(\widehat{2}, 3, \ldots,n,\widehat{P}_{n+1} )
= F(1,2,\ldots,n+1)\,,
\end{equation}
where
\begin{equation}
\widehat{P}_{n+1} = p_1 + p_{n+1} + z_{n+1} \lambda_2 \widetilde{\lambda}_1,
\qquad z_{n+1} = - \frac{ \ket{1}{n+1} }{\ket{2}{n+1}}\,.
\end{equation}
Then we use~(\ref{newmhv}) to write the $n+1$ graviton amplitude as
\begin{align}
\label{aaa}
{\mathcal{M}}_{n+1} &=
\sum_{k=4}^{n+1} \frac{\ket{1}{2} \ket{3}{k}}{\ket{1}{3}
\ket{2}{k}} M_k\cr
&= \frac{1}{(n-2)!} \sum_{\mathcal{P}(4,\ldots,n+1)}
\sum_{k=4}^{n+1} \frac{\ket{1}{2} \ket{3}{k}}{\ket{1}{3}
\ket{2}{k}} M_k\cr
&= \frac{1}{(n-3)!} \sum_{\mathcal{P}(4,\ldots,n+1)}
\frac{\ket{1}{2} \ket{3}{n+1}}{\ket{1}{3}
\ket{2}{n+1}} M_{n+1}\,.
\end{align}
On the second line we have used the fact that ${\mathcal{M}}_{n+1}$
is fully symmetric under the exchange of any labels to introduce
a sum over permutations together with $1/(n-2)!$ to compensate for
overcounting.  Inside the sum over permutations we can then without
loss of generality set $k = n+1$ while including a factor of $n-2$
counting the number of terms in the sum.
Now according to the definition~(\ref{Mdef}) we have
\begin{equation}
M_{n+1}
= \int d^8 \eta\ {\mathcal{M}}_{\text{L}}(\widehat{1},n+1,-\widehat{P}_{n+1})
\frac{1}{(p_1+p_{n+1})^2} {\mathcal{M}}_{\text{R}}
(\widehat{2},3,\ldots,n,\widehat{P}_{n+1})\,.
\end{equation}
Plugging in~(\ref{EF2}) for ${\mathcal{M}}_R$ we find that the
extra factor in front goes along for the ride as we apply the
recursion~(\ref{FBCF}), leading to
\begin{equation}
M_{n+1} = \sum_{\mathcal{P}(4,\ldots,n)}
\frac{\ket{\widehat{P}_{n+1}}{2} \ket{3}{4}}{\ket{\widehat{P}_{n+1}}{3}
\ket{2}{4}} F(1,2,\ldots,n+1)\,.
\end{equation}
Substituting this into~(\ref{aaa}) we find that the redundant
inner sum over $\mathcal{P}(4,\ldots,n)$ cancels the $1/(n-3)!$ factor leading
to
\begin{equation}
{\mathcal{M}}_{n+1} = \sum_{\mathcal{P}(4,\ldots,n+1)}
\frac{ \ket{1}{2} \ket{3}{n+1}} { \ket{1}{3} \ket{2}{n+1}}
\frac{\ket{\widehat{P}_{n+1}}{2} \ket{3}{4}}{\ket{\widehat{P}_{n+1}}{3}
\ket{2}{4}} F(1,2,\ldots,n+1)\,.
\end{equation}
Finally an elementary manipulation reveals that
\begin{equation}
\label{example1}
\frac{ \ket{1}{2} \ket{3}{n+1}} { \ket{1}{3} \ket{2}{n+1}}
\frac{\ket{\widehat{P}_{n+1}}{2} \ket{3}{4}}{\ket{\widehat{P}_{n+1}}{3}
\ket{2}{4}} = 
\frac{\ket{1}{2} \ket{3}{4}}{\ket{1}{3} \ket{2}{4}}\,,
\end{equation}
thereby completing the inductive proof of~(\ref{EF2}).
Note that we could have used a version of~(\ref{newmhv}) to delete
any one subamplitude of our choice, not necessarily $M_3$, so a byproduct
of our analysis is a demonstration that~(\ref{EF2}) is not
dependent on the choice of the distinguished particle $3$.
This feature was only checked
numerically in~\cite{Elvang:2007sg}.

\subsection{A new formula based on one
by Bedford, Brandhuber, Spence and Travaglini}

Here we apply the same idea to another $(n-2)!$-type formula, given
in (1.5) of~\cite{Bedford:2005yy} and reproduced here in a slightly relabeled
form,
\begin{equation}
\label{BBST}
{\mathcal{M}}_n =\frac{1}{2} \sum_{P(3,\dots,n)}
\frac{\bra{1}{n}}{\ket{1}{n} \ket{1}{2}^2}
\frac{\bra{3}{4}}{\ket{2}{3} \ket{2}{4} \ket{3}{4}
\ket{3}{5} \ket{4}{5}}
\prod_{s=5}^{n-1} \frac{\langle 2
| p_3 + p_4 + \cdots + p_{s-1} | s]}{\ket{s}{s+1} \ket{2}{{s+1}}}\,.
\end{equation}
Note that this formula is only valid for $n \ge 5$ (and for $n=5$ one simply
omits the product).
It was proven in~\cite{Elvang:2007sg} that this formula is equivalent
to~(\ref{EF1}).

Following the example set in the previous subsection it is clear that
we can simplify this formula by omitting one particle, say $3$,
from the sum over permutations at the expense of
introducing an appropriate factor like in~(\ref{EF2}).
Here we should be careful though because the formula~(\ref{BBST})
starts only from $n=5$
points so we should not choose the factor to involve particle 4, but rather
the factor must be
\begin{equation}
\label{extra}
1-\frac{z_5}{z_3}= \frac{ \ket{1}{2} \ket{3}{5} } { \ket{1}{3} \ket{2}{5} }\,.
\end{equation}
Inserting this factor into~(\ref{BBST})
leads to the simplified formula
\begin{equation}
\label{BBST2}
{\mathcal{M}}_n =\frac{1}{2} \sum_{P(4,\dots,n)}
\frac{\bra{1}{n}}{\ket{1}{n}}
\frac{\bra{3}{4}}
{\ket{1}{2} \ket{1}{3} \ket{2}{3} \ket{2}{4} \ket{2}{5} \ket{3}{4}
\ket{4}{5}}
\prod_{s=5}^{n-1} \frac{\langle 2
| p_3 + p_4 + \cdots + p_{s-1} | s]}{\ket{s}{s+1} \ket{2}{{s+1}}}\,.
\end{equation}
We have checked numerically that this modified version of~(\ref{BBST})
agrees with all of the other MHV formulas ((\ref{BGK}), (\ref{EF1}),
(\ref{EF2}) and~(\ref{MS})) through $n=12$ gravitons.
Of course it is also simple to prove analytically that it is correct,
following exactly the same steps as in the previous subsection.
One ends up with the same factors as in~(\ref{example1}) except
with $4 \to 5$, thereby establishing that~(\ref{BBST2}) is correct.

\subsection{Alternate proof of a formula due to Mason and Skinner}

Our final application concerns a formula established
recently by Mason and Skinner~\cite{Mason:2008ms}
\begin{equation}
\label{MS}
{\mathcal{M}}_n = \sum_{\mathcal{P}(2,\ldots,n-2)}
\frac{A(1,2,\ldots,n)}{\ket{1}{n-1} \ket{n-1}{n} \ket{n}{1}}
\prod_{m=2}^{n-1}
\frac{[m| p_{k+1} + p_{k+2} + \cdots + p_{n-1}|n\rangle}{\ket{m}{n}}\,.
\end{equation}
via a background field calculation of a single
graviton scattering off a self-dual gravitational background.
It was also shown to be equivalent to the original BGK formula~(\ref{BGK}),
thereby providing the first analytic proof of the BGK conjecture.

\begin{figure}
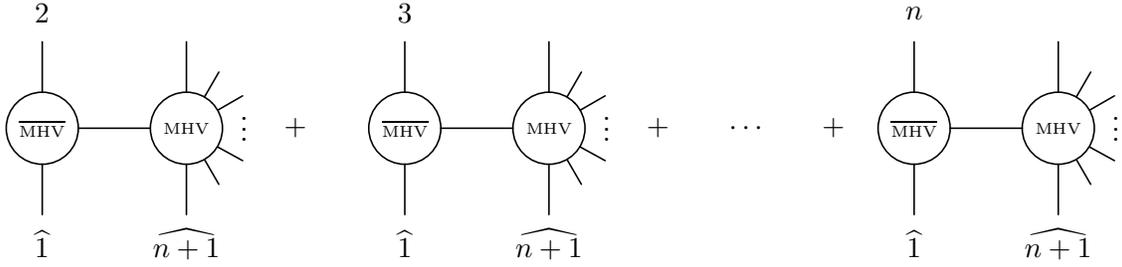

$$
{\hbox{\lower 57.5pt\hbox{
\begin{feynartspicture}(120,120)(1,1)
\FADiagram{} \FAProp(5.,7.5)(5.,7.5)(5.,12.5){/Straight}{0}
\FAProp(16.7678,11.7678)(16.7678,11.7678)(13.2322,8.2322){/Straight}{0}
\FAProp(7.5,10.)(12.5,10.)(0.,){/Straight}{0}
\FAProp(5.,7.5)(5.,4.)(0.,){/Straight}{0}
\FAProp(15.,7.5)(15.,4.)(0.,){/Straight}{0}
\FAProp(5.,12.5)(5.,16.)(0.,){/Straight}{0}
\FAProp(15.,12.5)(15.,16.)(0.,){/Straight}{0}
\FALabel(5.,2.)[]{$\widehat{1}$}
\FALabel(5.,18.)[]{$2$} \FALabel(15.,2.)[]{$\widehat{n+1}$}
\FALabel(5.,10.)[]{$\overline{\text{\tiny MHV}}$} \FALabel(15.,10.)[]{\tiny MHV}
\FALabel(19.,10.5)[]{$\vdots$}
\FAProp(17.1651,11.25)(18.8971,12.25)(0.,){/Straight}{0}
\FAProp(16.25, 12.1651)(17.25, 13.8971)(0.,){/Straight}{0}
\FAProp(16.25, 7.8349)(17.25, 6.1029)(0.,){/Straight}{0}
\FAProp(17.1651, 8.7)(18.8971, 7.75)(0.,){/Straight}{0}
\end{feynartspicture}
}}} + {\hbox{\lower 57.5pt\hbox{
\begin{feynartspicture}(120,120)(1,1)
\FADiagram{} \FAProp(5.,7.5)(5.,7.5)(5.,12.5){/Straight}{0}
\FAProp(16.7678,11.7678)(16.7678,11.7678)(13.2322,8.2322){/Straight}{0}
\FAProp(7.5,10.)(12.5,10.)(0.,){/Straight}{0}
\FAProp(5.,7.5)(5.,4.)(0.,){/Straight}{0}
\FAProp(15.,7.5)(15.,4.)(0.,){/Straight}{0}
\FAProp(5.,12.5)(5.,16.)(0.,){/Straight}{0}
\FAProp(15.,12.5)(15.,16.)(0.,){/Straight}{0}
\FALabel(5.,2.)[]{$\widehat{1}$} \FALabel(5.,18.)[]{$3$}
\FALabel(15.,2.)[]{$\widehat{{n+1}}$}
\FALabel(5.,10.)[]{$\overline{\text{\tiny MHV}}$} \FALabel(15.,10.)[]{\tiny MHV}
\FALabel(19.,10.5)[]{$\vdots$}
\FAProp(17.1651,11.25)(18.8971,12.25)(0.,){/Straight}{0}
\FAProp(16.25, 12.1651)(17.25, 13.8971)(0.,){/Straight}{0}
\FAProp(16.25, 7.8349)(17.25, 6.1029)(0.,){/Straight}{0}
\FAProp(17.1651, 8.7)(18.8971, 7.75)(0.,){/Straight}{0}
\end{feynartspicture}
}}} + \qquad \cdots\qquad +  {\hbox{\lower 57.5pt\hbox{
\begin{feynartspicture}(120,120)(1,1)
\FADiagram{}
\FAProp(3.,7.5)(3.,7.5)(3.,12.5){/Straight}{0}
\FAProp(14.7678,11.7678)(14.7678,11.7678)(11.2322,8.2322){/Straight}{0}
\FAProp(5.5,10.)(10.5,10.)(0.,){/Straight}{0}
\FAProp(3.,7.5)(3.,4.)(0.,){/Straight}{0}
\FAProp(13.,7.5)(13.,4.)(0.,){/Straight}{0}
\FAProp(3.,12.5)(3.,16.)(0.,){/Straight}{0}
\FAProp(13.,12.5)(13.,16.)(0.,){/Straight}{0}
\FALabel(3.,2.)[]{$\widehat{1}$}
\FALabel(3.,18.)[]{$n$}
\FALabel(13.,2.)[]{$\widehat{n+1}$}
\FALabel(3.,10.)[]{$\overline{\text{\tiny MHV}}$}
\FALabel(13.,10.)[]{\tiny MHV}
\FALabel(17.,10.5)[]{$\vdots$}
\FAProp(15.1651,11.25)(16.8971,12.25)(0.,){/Straight}{0}
\FAProp(14.25, 12.1651)(15.25, 13.8971)(0.,){/Straight}{0}
\FAProp(14.25, 7.8349)(15.25, 6.1029)(0.,){/Straight}{0}
\FAProp(15.1651, 8.7)(16.8971, 7.75)(0.,){/Straight}{0}
\end{feynartspicture}
}}}
$$\vskip -0.5cm \caption{Here we choose to use the
bonus relation to delete the last subamplitude.}
\label{mhvfig4}
\end{figure}

Here we provide an alternate proof of this formula by showing directly
that it satisfies the simplified on-shell recursion~(\ref{newmhv}).
As usual the proof proceeds by induction, beginning with the case $n=4$
which is simple to verify explicitly.  We then assume that~(\ref{MS})
holds for $n$ gravitons and apply the recursion~(\ref{newmhv}) to calculate
the $n+1$-graviton amplitude.
Choosing now for convenience
to shift $\lambda_1$ and
$\widetilde{\lambda}_{n+1}$ instead of~(\ref{bcfshift})
leads to the $n-1$ factorizations shown
in~\fig{mhvfig4}.
Furthermore we use the bonus relation
to delete the last subamplitude, so that the $k$-th
subamplitude picks up the factor
\begin{equation}
\label{newchoice}
1 - \frac{z_k}{z_n} = \frac{\ket{1}{n+1} \ket{n}{k}}
{\ket{1}{n} \ket{n+1}{k} }\,.
\end{equation}

The combinatorics work out just as in section III.C, so that
inside the appropriate sum over permutations we can
focus without loss of generality on just the first subamplitude
$k=2$ in~\fig{mhvfig4}.
Therefore let us now take a look at the factors which appear when~(\ref{MS})
is inserted into this diagram,
\begin{align}
&\frac{\ket{1}{n+1} \ket{n}{2}}
{\ket{1}{n} \ket{n+1}{2} } \times
\left[
\frac{1}{ \bra{2}{\widehat{P}_2}  \bra{\widehat{P}_2}{1}
\bra{1}{2} } \right]^2
\times \frac{1}{(p_1 + p_2)^2}\cr
&\qquad\times 
\frac{ A(\widehat{P}_2,3,\ldots,\widehat{n+1})}{
\ket{\widehat{P}_2}{n} \ket{n}{n+1} \ket{n+1}{\widehat{P}_2} }
\prod_{m=3}^n \frac{
[m|p_{m+1}+p_{m+2} + \cdots + p_n|n+1\rangle}{\ket{m}{n+1}}\,.
\label{missing}
\end{align}
The first term comes from~(\ref{newchoice}), the second term
is the 3-particle $\overline{\text{MHV}}$ amplitude on the left side
of the factorization, the third term is the propagator, and the
second line comes from inserting the $n$-graviton amplitude~(\ref{MS}) on
the right side of the factorization.
Now the factors
\begin{equation}
\frac{1}{ \bra{2}{\widehat{P}_2}  \bra{\widehat{P}_2}{1}
\bra{1}{2} } \times \frac{1}{(p_1 + p_2)^2} \times
A(\widehat{P}_2,3,\ldots,\widehat{n+1})
\end{equation}
are precisely those that would appear in the recursion for the MHV
SYM amplitude; these therefore
combine to give $A(1,2,\ldots,n+1)$.
Next we take a look at the factors
\begin{align}
&
\frac{\ket{1}{n+1} \ket{n}{2}}
{\ket{1}{n} \ket{n+1}{2} } \times
\frac{1}{ \bra{2}{\widehat{P}_2}  \bra{\widehat{P}_2}{1}
\bra{1}{2} } \times
\frac{1}{\ket{\widehat{P}_2}{n} \ket{n}{n+1} \ket{n+1}{\widehat{P}_2} }
\cr
&\qquad =
\frac{1}{\ket{1}{n} \ket{n}{n+1} \ket{n+1}{1}}
\times \frac{[2|p_1|n+1\rangle}{\ket{n+1}{2}}\,.
\end{align}
The second term here exactly
supplies the missing $m=2$ term in the product on the second
line of~(\ref{missing}),
so that when everything is finally combined
we end up with
\begin{equation}
\frac{A(1,2,\ldots,n+1)}{\ket{1}{n} \ket{n}{n+1} \ket{n+1}{1}}
\prod_{m=2}^n \frac{
[m|p_{m+1}+p_{m+2} + \cdots + p_n|n+1\rangle}{\ket{m}{n+1}}\,,
\end{equation}
in precise agreement with~(\ref{MS}), thereby completing the inductive proof.

\section*{Acknowledgments}

We are grateful to Nima Arkani-Hamed,
Zvi Bern, Lance Dixon, James Drummond,
Henriette Elvang,
Dan Freedman, and
Steve Naculich for stimulating correspondence and discussions, and to
Chrysostomos Kalousios and
Cristian Vergu for collaboration in the early stages of this work.
This work was supported in part by the US
Department of Energy under contract DE-FG02-91ER40688 (MS (OJI)
and AV), and the US National Science Foundation under grants
PHY-0638520 (MS) and PHY-0643150 CAREER and PECASE (AV).

\end{document}